\documentclass[twocolumn,english]{article}
\usepackage{authblk}
\usepackage{etoolbox}
\usepackage{lmodern}
\usepackage{subcaption}
\usepackage{graphicx}
\usepackage[utf8]{inputenc}

\makeatletter 
\def\@normalsize{\@setsize\normalsize{10pt}\xpt\@xpt
\abovedisplayskip 10pt plus2pt minus5pt\belowdisplayskip \abovedisplayskip
\abovedisplayshortskip \z@ plus3pt\belowdisplayshortskip 
6pt plus3pt minus3pt\let\@listi\@listI}
\def\subsize{\@setsize\subsize{12pt}\xipt\@xipt}
\def\section{\@startsection {section}{1}{\z@}
	{1.8ex plus 1ex minus .2ex} 
	{1.2ex plus .2ex \@afterindentfalse}
	{\large\bf}}
\def\subsection{\@startsection {subsection}{2}{\z@}
	{1.3ex plus 1ex}
	{.8ex plus .2ex \@afterindentfalse}
	{\subsize\bf}}
\def\paragraph{\@startsection {paragraph}{4}{\z@}
	{1.8ex plus .3ex}
	{-1em  \@afterindentfalse}
	{\normalsize\bf}}

\let\OLDthebibliography\thebibliography
\renewcommand\thebibliography[1]{
	\OLDthebibliography{#1}
	\setlength{\parskip}{0pt}
	\setlength{\itemsep}{0pt plus 0.3ex}
}

\patchcmd{\@maketitle}{\LARGE \@title}{\fontsize{16}{19.2}\selectfont\@title}{}{}
\makeatother

\title{Tool-Supported Experiments for Continuously Collecting Data of Subjective Video Quality Assessments During Video Playback}

\author{Oliver Karras}
\author{Jil Klünder}
\author{Kurt Schneider}
\affil{Leibniz Universität Hannover, Software Engineering Group, 30167 Hannover, Germany}
\affil{Email: \{oliver.karras, jil.kluender, kurt.schneider\}@inf.uni-hannover.de}
\date{}

\begin{document}
\maketitle
\section{Introduction}
The adequate use of documentation for communication is one challenge in requirements engineering (RE) \cite{Karras.2018b}. In recent years, several researchers \cite{Fricker.2016, Karras.2017, Rupp.2018} addressed this challenge by using videos as a communication mechanism. All of them concluded that this way of using videos has the potential to facilitate requirements communication. Nevertheless, software professionals are not directors and thus do not necessarily know what constitutes a good video \cite{Karras.2018}. This lack of knowledge is one crucial reason why videos are still not an established communication mechanism in RE \cite{Karras.2018, Rupp.2018}. When videos shall be established in the RE activities, practices, and techniques, requirements engineers have to acquire the necessary knowledge to produce and use good videos on their own at moderate costs, yet sufficient quality.

In our research project \textit{ViViReq} (see Acknowledgment), we aspire to bridge this knowledge gap about what constitutes a good video. Whether a video is good or not depends on its quality perceived by its viewers \cite{Karras.2018}. However, video quality is a rather ill-defined concept due to numerous unspecified technical and subjective characteristics \cite{Winkler.2008}. As part of our research plan, we develop a quality model for videos \cite{Karras.2019} inspired by the idea of Femmer and Vogelsang \cite{Femmer.2018} to define and evaluate the quality of videos as RE artifacts. In addition to evaluating videos, this quality model can be used to identify the relevant characteristics of videos for their specific purpose which can be further used to specify requirements, their criteria for satisfaction, and corresponding measures. Therefore, software professionals may use the quality model as guidance for producing and using videos.

\section{Background and Problem}
Based on our quality model \cite{Karras.2019}, we want to analyze how specific implementations of single video characteristics, e.g., pleasure, exactly affect the viewers' quality assessment. Subjective video quality assessment (VQA) is the most accurate and reliable practice to determine video quality \cite{Akramullah.2014}. Experiments are an established way for subjective VQAs since a controlled environment is necessary to avoid potential disruptive factors, e.g., varying viewing conditions \cite{Akramullah.2014}. In these experiments, $15$ to $30$ subjects individually watch one or more videos and assess the quality of each video directly after the video was completely watched. Each subject assesses his perception of the video quality by assigning a single value on a defined $5$- or $7$-level scale to each video. For each video, the average of the subjects' assessments is calculated. This value is defined as the Mean Opinion Score (MOS) which indicates the perceived overall quality of the respective video from the subjects' point of view \cite{Winkler.2008}.

The established subjective VQA methods (cf. \cite{Akramullah.2014}) ultimately determine video quality based on the MOS. Therefore, these methods only relate to an entire video, so that it is not possible to determine exactly which particular parts of a video had a particularly positive or negative impact on a subject. Thus, the established subjective VQA methods do not lend themselves to detailed analyses of videos to determine how specific implementations of single video characteristics exactly affect the viewers' quality assessment.

\section{Solution: A Research Preview}
\begin{figure*}[!t]
	\captionsetup{justification=centering}
	\begin{subfigure}{.325\textwidth}
		\centering
		\includegraphics[width=\columnwidth]{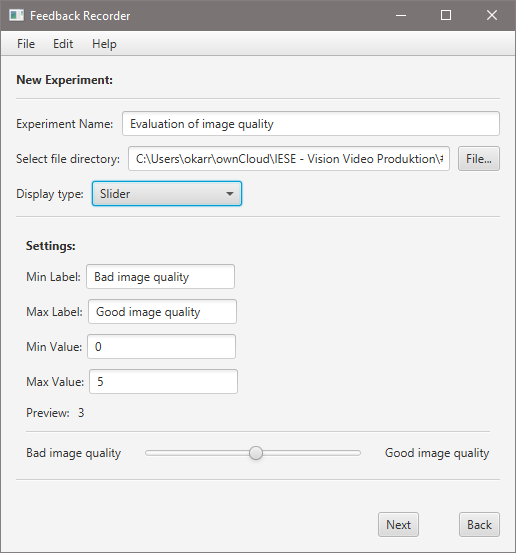}
		\caption{Customizable input method}
		\label{sfig1}
	\end{subfigure}
	\begin{subfigure}{.325\textwidth}
		\centering
		\includegraphics[width=\columnwidth]{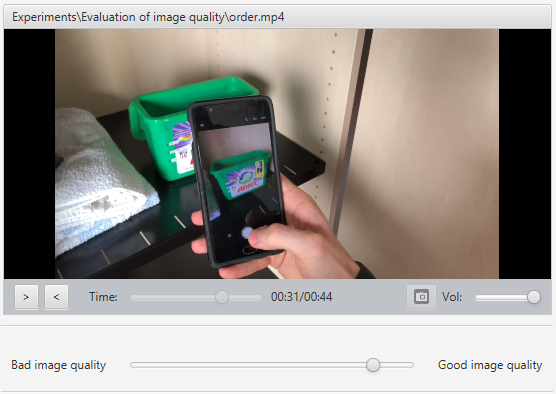}
		\caption{Continuous data collection}
		\label{sfig2}
	\end{subfigure}
	\begin{subfigure}{.325\textwidth}
		\centering
		\includegraphics[width=\columnwidth]{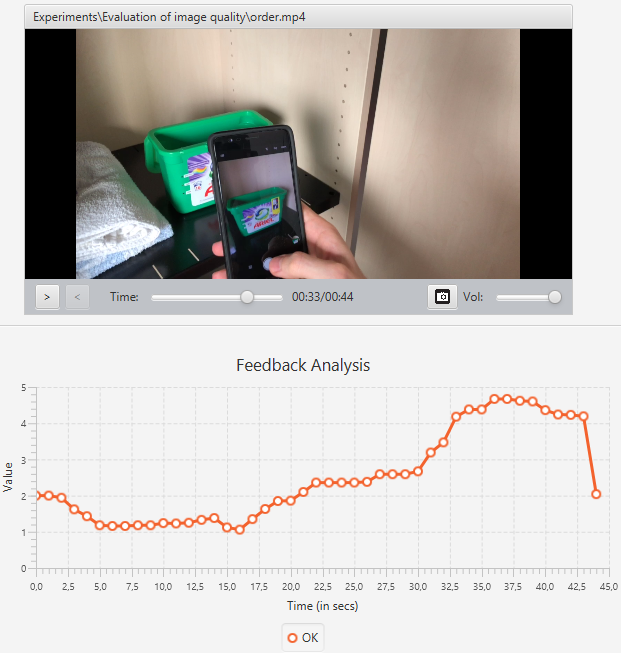}
		\caption{Immediate data visualization}
		\label{sfig3}
	\end{subfigure}
	\centering
	\caption{Screenshots of three concepts of the prototype}
	\label{fig}
\end{figure*}

Considering the established subjective VQA methods, our intended detailed analysis of videos is not possible. Thus, we propose to support the conduct of experiments for subjective VQAs with a software tool which allows continuously collecting the assessment data for generic video characteristics during video playback. The proposed solution consists of five key concepts which are briefly explained below. \figurename{ \ref{fig}} shows the implementation of three concepts in our prototype.

\paragraph{Management of Experiments.}
The software tool supports the management of several experiments and their associated subjects to make it easier for the experimenter to handle the respective data sets.

\paragraph{Customizable Input Method.}
\figurename{ \ref{sfig1}} shows the creation of an experiment. Each experiment includes a name, the video to be examined, and a customizable input method for the continuous data collection during video playback. At the moment, the software tool supports a slider as well as two up to ten radio buttons as an input method. These input methods are completely customizable concerning the labels, the scale, and the number of radio buttons.

\paragraph{Continuous Data Collection.}
\figurename{ \ref{sfig2}} presents the continuous data collection. In this case, a subject can continuously assess his current perception of the quality characteristic to be examined during video playback by simply moving the slider.

\paragraph{Immediate Data Visualization.}
After data collection, the experimenter can examine the collected data of all subjects together and each subject individually (see \figurename{ \ref{sfig3}}). In this way, an experimenter can get an overview of the collected data and analyze it immediately with the subject. Therefore, the experimenter can gain deeper insights into what particular parts of the video affected the subject's assessments.

\paragraph{Data Export.}
An experimenter can export all collected data as CSV files for later analysis.\\

\noindent
The results of a first evaluation indicated that the subjects did not feel distracted by the continuous data collection and preferred the slider as input method.

\section{Conclusion}
Already a single very good or very bad implemented characteristic of a RE artifact, e.g., a video, can significantly affect its overall assessment by a stakeholder. In this paper, we propose five key concepts for a software tool that is intended to enable detailed analyses of videos which are currently not supported by established subjective VQA methods. 

First, we assume that the software tool may be beneficial for other researchers and practitioners who also deal with videos and their assessments due to its flexibility. Second, our intended use of such a detailed analysis of videos is to provide fine-grained insights into the interrelationships of the specific implementation of a video characteristic and its impact on the viewers' quality assessment. Based on these insights, we expect to be able to derive recommendations about what characteristics constitute a good video and how these characteristics should be implemented in a video. These recommendations are important to operationalize the quality model. They enhance the practical utility of the quality model by revealing how the single video characteristics can be useful in practice. Based on our quality model and the derived recommendations, we suppose to provide software professionals with the knowledge necessary to produce and use good videos in RE on their own at moderate costs and with sufficient quality.

\section*{Acknowledgment}\label{sec:acknowledgment}
This work was supported by the Deutsche Forschungsgemeinschaft (DFG) under Grant No.: $289386339$, ($2017$ -- $2019$).

\bibliographystyle{abbrvdin} 
{\scriptsize \bibliography{references}}
\end{document}